\documentclass[fleqn]{mnras}

\renewcommand{\footnoterule}{%
  \kern -19pt
  \hrule width 2in
  \kern 2.6pt
}
\usepackage{mathptmx}
\usepackage[T1]{fontenc}
\usepackage{ae,aecompl}
\usepackage{lscape}
\usepackage{epstopdf}
\usepackage{epsfig}
\usepackage{graphics}
\usepackage{graphicx}
\usepackage{longtable}
\usepackage{amsmath}	    
\usepackage{amssymb}	
\usepackage{scrextend}  
\usepackage{xcolor}         
\usepackage{breqn}
\usepackage{gensymb}     

\title[On Alignment of  Galaxies in Clusters]{On Alignment of  Galaxies in Clusters}

\author[Hrant~M.~Tovmassian]{
Hrant~M.~Tovmassian,$^{1}$\thanks{E-mail: htovmas@gmail.com (HMT)}
and J.P. Torres-Papaqui$^{2}$\thanks{E-mail: jp.torrespapaqui@ugto.mx (JPTP)}
\\
$^{1}$377, W. California 30, Glendale, CA, USA\\
$^{2}$Departamento de Astronom\'ia, Universidad de Guanajuato, Apartado Postal 144, 36000, Guanajuato, Mexico\\
}

\date{Accepted XXX. Received YYY; in original form ZZZ}

\pubyear{2019}

\begin{document}
\label{firstpage}
\pagerange{\pageref{firstpage}--\pageref{lastpage}}
\maketitle

\begin{abstract}
We discover alignment of galaxies in clusters by analyzing the distribution of their position angles. 
We assume that galaxies are aligned, if their number at one $90\deg$ position angle interval is more 
than twice higher than at another $90\deg$ interval. We selected for study the isolated clusters in 
order to exclude clusters the distribution of the galaxy orientation in which could possibly be 
altered by gravitational influence of a nearby cluster. Also, we study the galaxy position angle 
distribution at the outer regions of clusters with small space density where the variation of the 
position angles in the result of interactions between galaxies is smaller than at the central dens 
regions used as a control sample. We found that the alignment of galaxies is more often observed in 
poor clusters. We conclude that originally galaxies were aligned, but  in the result of accretion of 
field galaxies with arbitrary orientation and mutual interactions of galaxies during the cluster 
evolution the relative part of aligned galaxies decreases.
\end{abstract}


\section{Introduction}

The alignment of galaxies in clusters could be clue to the origin of clusters. In the pancake 
scenario (Zel'dovich, 1970; Zel'dovich et al. 1982; Doroshkevich, Shandarin, \& Saar 1978) 
galaxies form after the gas-dust cloud collapse and the position angles (PA) of galaxies will 
be aligned. In the past some evidence on the alignment of galaxies with the parent cluster were 
reported by Adams, Strom \& Strom (1980), Sastry (1968), Carter \& Metcalfe (1980), Binggeli (1982), 
Struble \& Peebles (1985), Rhee \& Katgert (1987), Lambas et al. (1988), Flin \& Olowin (1991), 
Fong, Stevenson, \& Shanks (1990. More certainly the alignment were found between orientations 
of the cluster and of the BCG (cD) (Sastry 1968; Binggeli 1982; Carter \& Metcalfe 1980; 
Struble 1990; West 1989; West 1994; Plionis 1994; Fller, West \& Bridges 1999; Kim et al. 2001; 
Chambers, Melott, Miller 2002). Plionis et al. (2003) and Rong,  Zhang \& Liao (2015) found an 
evidence that significant galaxy alignments are present in dynamically young clusters. 
Meanwhile, Dekel (1985), van Kampen \& Rhee (1990), Trevese, Cirimele \& Flin (1992), 
Djorgovski 1987 and Cabanela \& Aldering (1998) found no galaxy alignment, except the alignment 
of the BCG with its parent cluster.  

Hence, no a certain conclusion was achieved on the galaxy alignment in clusters. In this paper 
we undertook new search for alignment of galaxies in clusters and showed that galaxy alignment 
is mostly revealed in poor clusters. It is suggested that at the cluster origin the primordial 
orientations of member galaxies are ordered and that assembly of field galaxies later on and 
interactions between galaxies introduce disorder in the the galaxy orientations.

\section {The data}

We study the possible alignment of galaxies in ACO  (Abell, Corwin, \& Olowin, 1989) clusters.
Many Abell clusters are themselves clustered (Abell 1958; Bogart \& Wagoner 1973; Hauser 
\& Peebles 1973). The gravitational influence of the nearby cluster may affect on the 
orientation of galaxies in the studied cluster. In order to exclude this effect we selected for 
study isolated clusters. We assume that the cluster is isolated, if the nearest neighbor ACO 
cluster is located at the projected distance $>10$~Mpc on the sky. 

Note that any primordial galaxy alignments in clusters could be severely damped by 
the violent relaxation, by the exchange of angular momentum in galaxy interactions that 
occur in the dense cluster environment over a Hubble time (Coutts 1996) and also by 
secondary infall (Quinn \& Binney 1992). Therefore, we searched the orientation of galaxies in 
the outer area of clusters. The member galaxies of the most ACO clusters are located within 
2 Mpc of the Abell radius (1958), defined as $R_A$ = 2~Mpc $h_{72}^{-1}$ = 1.7'/z $R_A$ = 
2 Mpc $h_1= 1.7$/z 72 (Andernach, Waldhausen,  \& Wielebinski 1980) where z is the cluster 
redshift. In the selected clusters we studied the distribution of position angles (PAs) of 
galaxies in the ring with cluster-centric radii $1\div2$ Mpc. As a control sample we use the 
central area (further - the control area) with 1 Mpc radius of the same clusters. The clusters 
at $z<0.1$ with more than 10 galaxies within the the studied region and the control area were 
studied. The used parameters of clusters are taken from NED. Finally for the alignment search 
we selected 75 isolated ACO clusters. 
 
The member galaxies in the studied clusters were retrieved from photometric catalog of the 
Ninth Data Release (DR9) of the Sloan Digital Sky Server (SDSS) project (Ahn et al. 2012). 
The retrieved galaxies are those with the primary mode (marked by 1) and good quality of 
observations (marked by 3). Also, the galaxies with velocities within $\pm1500$ km s$^{-1}$ 
of the cluster velocity (Collins et al. 1995) were retrieved.\footnote {The NASA/IPAC Extragalactic 
Database (NED) is operated by the Jet Propulsion Laboratory, California Institute of Technology, 
under contract with the National Aeronautics and Space Administration.}

\section {Analysis}

For search of the alignment of the orientation of galaxies in clusters we divided the range of PAs 
of galaxies in each cluster into two $90\deg$ sections so that to have high number ($N_h$)  of 
galaxies at one section and small number ($N_s$) of galaxies at the other section. We assume 
that there is an alignment signal, if the number of galaxies at one $90\deg$ section is by at least 2 
times higher than at the other $90\deg$ section. The results of counts are presented in Table \ref{Table01}.

At corresponding columns of Table~\ref{Table01} the following information is given: 1st - the 
cluster designation; 2d - the redshift of the cluster; 3d - the number of galaxies in the searched 
region, 4d - the range of PAs in $90\deg$ section with high number of galaxies; 5th - the 
interval of PAs at which the high number of galaxies are distributed;  6th - the number $N_h$ of 
galaxies at this section; 7th - the number of galaxies at the opposite section; 8th - the ratio 
$N_h/N_s$ at the searched region; 9th - the ratio $N_h/N_s$ at the control region.

\begin{table*}
\caption{The results of counts of galaxies with PAs in two $90\deg$ intervals.}
\label{Table01}
\begin{tabular}{lllllllll}
\hline\hline
Cluster &   $z$  & $N_2$ & $90\deg$ & $I_{PA}$ &  $N_h$ & $N_s$ &  $\left(\frac{N_h}{N_s}\right)_{1\div 2}$ & $\left(\frac{N_h}{N_s}\right)_1$  \\
\hline
	  &        &       &  degree  & degree &         &   &  &  \\				
\hline
A595  & 0.0666 &   51  &  32-122 & 85 &  12 &   3  &    4.0  &  1.0   \\
A602  & 0.0619 &   62  &  69-159 & 89 &  20 & 10  &    2.0  &  1.7   \\
A634  & 0.0265 &  104  &  60-150 &   -  &  20 & 16  &   1.25 &  1.7  \\
A635  & 0.0925 &   34  &  78-165 & 87  &   8  &  2   &   4.0  &  1.4   \\  
A660  & 0.0642 &   26  & 163-73  & 88  &  12 &  3   &   4.0  &  3.5  \\   
A671  & 0.0502 &   99  &  79-169 &   -  &  25  & 14  &   1.5  &  1.5  \\
A690  & 0.0788 &   48  &  10-101 &   -   &  17 &   9  &   1.9  &  2.1   \\
A695  & 0.0687 &   27  &  26-116 & 90  &    8 &   2  &    4.0  &  2.4   \\  
A699  & 0.0851 &   31  &  68-158 & 80  &  12 &   6  &    2.0  &  1.6   \\  
A692  & 0.0894 &   48  &  19-109 &   -   &  15 & 10  &    1.5  &  1.6   \\
A724  & 0.0933 &   46  &  171-81 & 83  & 14 &   3  &   4.7 &  1.2   \\ 
A727  & 0.0951 &   57  &  148-58 &   -   &  24 & 15  &    1.6 &   5.0   \\
A744  & 0.0729 &   32  &  127-25 & 78  &    7 &   3 &    2.3 &    6.0  \\	
A757  & 0.0517 &   49  &  173-83 &   -   &  11 &   8  &   1.4 &   1.1  \\ 
A779  & 0.0225 &  112  &  15-105 &   -   &  23 &  17 &    1.4 &  1.1  \\
A819  & 0.0759 &   23  &  67-157 &  83 &    8 &    4 &    2.0 &  2.7  \\ 	
A834  & 0.0709 &   35  &  38-128 &  86 &  16 &    3 &    5.3 &  3.0   \\
A858  & 0.0863 &   26  &  46-136 &  80 &    8 &    2 &    4.0 &  1.3   \\ 
A1024 & 0.0734 &   49  &  15-105   & -     &  13 & 12  &    1.1 &  1.9   \\
A1028 & 0.0908 &   26  & 179-89  &  90 &  10 &   3  &    3.3 & 2.25  \\  
A1035 & 0.0684 &   57  & 69-159  &  90 &  10 &   2  &    5.0 & 1.6   \\
A1066 & 0.0690 &   81  & 75-165  &   -   &  18 &  13 &    1.4 & 1.4   \\
A1100 & 0.0463 &   53  &  90-180  &   -   &  15 &    8 &    1.9 &  2.0    \\
A1126 & 0.0646 &   33  & 18-108  &  88 &   11 &    5 &   2.2 & 16.1   \\
A1139 & 0.0398 &   50  &    6-96   &  80 &   12 &   6  &   2.0 & 1.5    \\  
A1142 & 0.0349 &   68  & 17-107  &   -   &   18 & 12  &   1.5 & 1.9    \\
A1168 & 0.0906 &   41  &   3-93	   & 81  &   12 &   5  &   2.4 &  2.7  \\  
A1169 & 0.0586 &   78  &   1-91    & 77  &   19 &   8  &   2.4 &  1.4   \\
A1238 & 0.0733 &   68  & 105-15  & 81  &   23 &   7  &   3.3 &  2.4  \\ 
A1270 & 0.0692 &   63  & 69-159  & 87  &   17 &   6  &   2.8 & 1.3   \\ 
A1307 & 0.0817 &   67  & 164-74  & 88  &   24 & 11  &   2.2 & 1.9   \\ 
A1314 & 0.0335 &  119  &  30-120 &   -   &   23 & 15  &   1.5 & 1.3   \\
A1371 & 0.0398 &   60  & 32-122  & 86  &   17 &   7  &   2.4 & 2.6  \\ 
A1424 & 0.0768 &   72  & 53-143  &  -    &   23 & 15  &   1.5 & 1.4 \\ 
A1480 & 0.0734 &   31  & 160-70  & 90  &   11 &   5  &    2.2 & 1.5   \\
A1507 & 0.0604 &   57  &    0-90   &   -   &  19  & 12  &   1.6  & 1.9  \\
A1516 & 0.0769 &   59  & 72-162  & 87  &   21 & 10  &   2.1  & 1.5   \\ 
A1541 & 0.0893 &   71  &   7-97   & 77   &  21  &  9   &   2.3  &  2.5 \\  
A1552 & 0.0858 &   75  & 151-61  & 90  &  24  & 12  &  2.0  &  2.1   \\ 
A1564 & 0.0792 &   54  &  32-122  &   -   &  16  & 12  &  1.3  &  2.25 \\
A1599 & 0.0855 &   25  & 100-10  & 68  &    7  &   3  &   2.3 &  2.0   \\ 
A1616 & 0.0833 &   75  &   0-90  &  87 &  19  &   8  &   2.4  &  1.6 \\   
A1630 & 0.0648 &   36  &  7-96   &  -    &   7   &   4  & 1.75  &  2.6  \\ 
A1658 & 0.0850 &   25  & 153-63  &   -   &   11 &   5  &  2.2  &  8.1   \\ 
A1684 & 0.0862 &   27  & 66-156  &   -   &    9  &   6  &  1.5  &  2.0   \\
A1692 & 0.0879 &   50  & 132-42  & 90  &  18  &   9  &   2.0 &  2.7   \\ 
A1750 & 0.0852 &   86  & 13-103  &   -   &  36  & 23  &  1.6  &  2.7   \\
A1781 & 0.0618 &   45  & 66-156  & 87  &  19  &   6  &   3.2  & 1.2   \\ 
A1783 & 0.0690 &   50  & 77-167  & 77  &  12  &   3  &   4.0  & 1.7   \\ 
A1809 & 0.0791 &   87  & 15-105  & 88  &   23 &    10  &   2.3 & 2.5   \\ 
A1812 & 0.0630 &   27  & 32-122  & 84  &  11  &   4  &   3.0  &  2.0  \\  
A1825 & 0.0595 &   30  & 110-20  & 90  &  11  &    5  &  2.2  &  2.5   \\ 
A1827 & 0.0654 &   40  & 78-168  & 77  &  12  &    6  &  2.0  & 1.75  \\
A1849 & 0.0963 &   27  & 73-163  & 77  &    9  &    2  &  4.5  &  3.0  \\  
A1864 & 0.0870 &   51  & 15-105  &  -    &  16  &  12  &  1.3  & 1.3   \\
A1890 & 0.0574 &   83  & 51-141  &  -    &   21 & 14  &  1.5  &  1.4    \\
A1939 & 0.0881 &   41  & 63-153  & 90  &   12 &   6  &   2.0  &  2.3   \\
\hline\hline	     				
\end{tabular}  									
\end{table*} 

\begin{table*}
\renewcommand\thetable{\ref{Table01}}
\caption{Continued.}
\begin{tabular}{lllllllll}
\hline\hline
Cluster &   $z$  & $N_2$ & $90\deg$ & $I_{PA}$ &  $N_h$ & $N_s$ &  
$\left(\frac{N_h}{N_s}\right)_{1\div 2}$ & $\left(\frac{N_h}{N_s}\right)_1$  \\
\hline
	       &             &             &  degree & degree &             &           &   &   \\				
\hline
A2018 & 0.0878 &   54 & 44-134  & 90  &   23 &   9  &   2.6  & 1.0  \\
A2019 & 0.0807 &   24 & 75-165  & 81  &     9 &   4  &  2.25 & 2.7   \\  
A2022 & 0.0578 &   77 &  6-96     &  -    &   21 & 15  &   1.4  &  1.4 \\
A2048 & 0.0972 &   60 &   0-90    &  -    &   19 & 14  &   1.4  &  2.0  \\
A2082 & 0.0862 &   24 & 130-40  & 87  &   11  &   1  &  11.0 &  5.0  \\
A2107 & 0.0411  & 130 & 15-105  &  -    &   33  & 22  &   1.5  &  1.3  \\
A2108 & 0.0919 &   50 & 32-122  & 76  &   13  &   4  &  3.25 & 2.3  \\  
A2110 & 0.0980 &   27 & 90-180  & 88  &   10  &   3  &    3.3 & 1.8   \\ 
A2122 & 0.0661 &   72 & 51-141  & 90  &   22  & 11  &    2.0 & 2.25 \\   
A2142 & 0.0909 & 123 & 88-178  &  -    &   36  & 26  &   1.4 & 1.25 \\
A2148 & 0.0877 &   29 & 108-18  &  90 &     7  &   3  &    2.3 & 1.1  \\ 
A2162 & 0.0322 &   46 & 98-180  &   -   &   10  &   6  &   1.7 &  1.5  \\
A2178 & 0.0928 &   46 & 171-81  & 83  &   14  &   3  &    4.7 & 1.2  \\ 
A2205 & 0.0876 &   39 & 46-136  &   -   &    13 &  11 &    1.4 & 1.5  \\ 
A2255 & 0.0806 & 122 & 16-106  &    -  &    28 &  26 &    1.1 & 1.4 \\
A2366 & 0.0529 &   56 & 89-179  & 56  &     9  &    4 &  2.25 & 1.7 \\ 
A2593 & 0.0413 & 138 &  2-92     &   -   &   36  &  24 &   1.5  & 1.1 \\
A2630 & 0.0667 &   36 & 40-130  & 85  &     9  &    2 &   4.5  & 1.5  \\ 
\hline\hline	     				
\end{tabular}  									
\end{table*} 
\addtocounter{table}{-1}

 In the case of a roughly uniform distribution of PAs the numbers of galaxies in two $90\deg$ 
intervals could occasionally differ from each other by more than 2 times. We found that in 
the ring with cluster-centric radii $1\div2$ Mpc of 47 out of the studied 75 clusters, i.e 
$\sim$63\% of them have alignment signal. The probability that in 47 clusters out of 75 the 
ratio $N_h/N_s$ exceeds 2 is 62.66\% of success and a 95 percent confidence interval from 50.73\% 
to 73.56\% with a p-value = 0.0369. Hence, the probability that the found alignments are not by 
chance is sufficiently high. Meanwhile at the control region the  alignment signal have 
32 clusters, $\sim$43\%. The probability that this is by chance is 42.66\% of success and a 95 
percent confidence interval from 31.30\% to 54.62\% with a p-value = 0.2480. Hence, the probability 
that the found alignments of galaxies at the cluster central region are caused by random distribution 
of PAs is high.

We more detailed  analyze the distribution of PAs in the outer region of clusters within radii 
$1\div2$ Mpc. We supposed that the alignment could depend on the richness of clusters and on 
the limiting magnitude of the observed galaxies. In order to find out whether the alignment signal 
depends on the cluster richness or distance we split the list of clusters into two parts  with poor and 
rich ones and also with high and low redshifts. The deduced parameters  are presented in 
Table \ref{Table02}. In the 1st column of Table \ref{Table02} the analyzed parameters are given: the 
redshift, the average number $N$ of galaxies, the average minimal absolute magnitude in r-band, 
the ratio of the number of clusters with alignment signal to the total number of galaxies, $N_{as}/N_t$ 
in the studied region, the average value of the ratio of the high number $N_h$ of galaxies in one 
$90\deg$ interval to the small number of galaxies at the other $90\deg$ interval for the outer ring. 
In the 2d and 3d columns the corresponding data related to the nearby and distant clusters are 
presented. 

Table \ref{Table02} shows that for nearby clusters at average $<z>=0.0559$ the relative number of 
clusters with alignment signal is 0.58 with 58.00\% of success and 95 percent confidence 
interval from 47.71\% to 67.80\% with a p-value = 2.182e-03. Meanwhile the relative number of  
distant clusters at average $<z>=0.0861$ with alignment signal is higher 0.70 with 70.00\% of 
success and a 95 percent confidence interval from 60.01\% to 78.75\% with a p-value = 9.818e-05.
The relative number of clusters with alignment signal in rich clusters at small average $<z>=0.0667$ 
is smaller of that of the poor clusters at higher average $<z>=0.0767$. The rich clusters with 
alignment signal has a 53.00\% of success and a 95 percent confidence interval from 42.75\% to 
63.05\% with a p-value = 6.173e-02, meanwhile poor clusters with alignment signal has a 77.00\% 
of success and a 95 percent confidence interval from 67.51\% to 84.82\% with a p-value = 5.514e-08.

Table  \ref{Table02} shows that the differences between the total numbers of galaxies, the limiting 
absolute magnitude, the ratios $N_h/N_s$ are higher for rich and poor clusters in comparison to 
those of nearby and distant clusters. The relative number of clusters with alignment signal in rich 
and poor cluster differs from each other by 1.6 times, while  this difference between nearby and 
distant clusters is smaller, 1.2. The difference between average absolute magnitudes $M_r$ between 
nearby and distant clusters is less than $1^m$ and is higher, $0.30^m$ between rich and poor clusters. 
The ratios $N_{as}/N_t$ and $N_h/N_s$ are also higher between clusters of different richness, 1.45 
and 1.2, and 1.38 and 1.04.

It follows that the defining reason for high relative number of clusters with alignment signal is not 
the presence of relatively large number of faint galaxies in them, but generally the cluster richness
The poorer they are the higher is the probability of preserving in them the primordial alignment of galaxies.  

\begin{table}
\caption{The parameters of nearby, distant, poor and rich clusters.}
\label{Table02}
\begin{tabular}{rrr}
\hline\hline
	& Nearby clusters & Distant clusters  \\
\hline
$<z>$		 & 0.0559  & 0.0861 \\
$<N>$		 &	62	&   50	\\
$<M_r>$		 & -19.11	& -20.04	\\
$N_{as}/N_t$ & 0.58	& 0.70	\\
$N_h/N_s$	 & 2.4	& 2.5	\\
\hline
	   &        Rich clusters  &   Poor clusters  \\				
\hline
$z$	          &  0.0666 & 0.0774   \\
$<N>$	      &    75	& 34		\\
$<M_r>$		  & -19.49	& -19.81	\\
$N_{as}/N_t$  &  0.53	& 0.77	\\
$N_h/N_s$	  &  2.1	& 2.9		\\ 
\hline\hline	     				
\end{tabular}  									
\end{table}

\begin{figure} 
\centering
\includegraphics[width=\columnwidth]{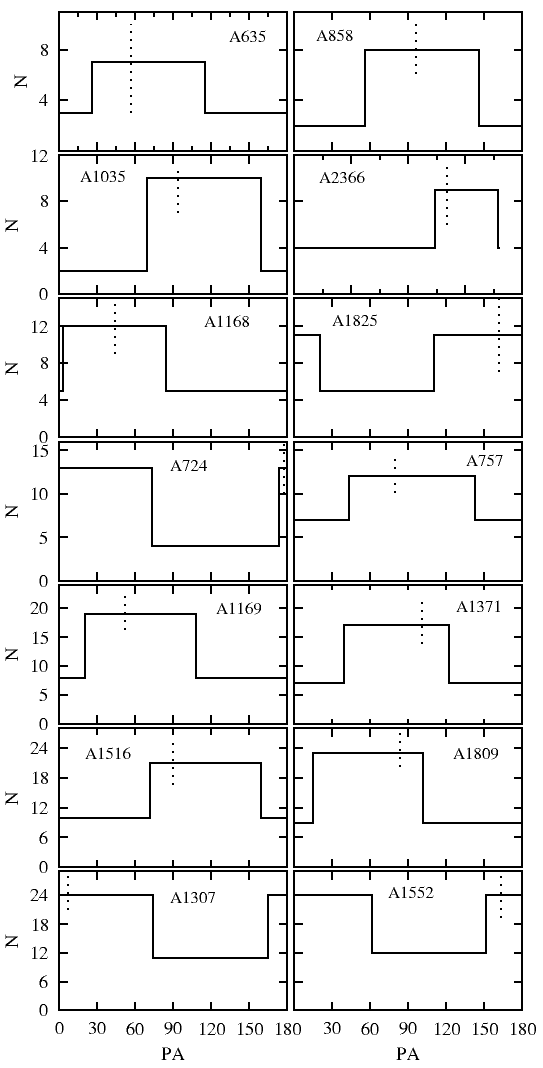}
\caption{The comparison of the distribution of PAs of galaxies in clusters with alignment signal with 
the cluster PA from Plionis (1994) shown by dotted line. }
\label{Figure01}
\end{figure}

The PAs of the large axes of 19  clusters of our list with alignment signal were determined by Plionis 
(1994).  In Figure~\ref{Figure01} the distribution of PAs of galaxies in 14 clusters and the PAs  of the 
cluster large axis are shown. In 14 clusters the PAs of clusters are within interval of PAs of aligned 
galaxies. The PAs of large axes of only 5 clusters, A1126, A1139, A1541, A1783 and A1812 are out 
of the interval of PAs of galaxies with alignment signal.The probability of 14 chance coincidences out 
of 19 is sufficiently small, 0.02. This proves that the applied method for searching the alignment is 
reliable. 

\section{Discussion and Conclusions}

By the study of the distribution of PAs of galaxies in the ring with radii $1\div2$ Mps of 75 isolated 
clusters the alignment signal is found in 47 clusters. Separate analysis of clusters of different richness 
showed that alignment depends on the cluster richness. The percentage of clusters with alignment is 
the highest, 78\%, in poor clusters containing on average 35 galaxies. In rich clusters with on 
average 76 galaxies the percentage of clusters with alignment signal is smaller, 50\%. The alignment 
signal is found also for galaxies of the control sample in the central area of clusters with 1 Mpc.
However, the percentages of clusters with alignment signal are smaller, 49\% for poor clusters and 
36\% for rich ones. Hence, the poorer is the cluster, the higher is the chance to reveal the alignment. 

The alignment of galaxies is found mostly in poor clusters. It evidences in favor of  pancake scenario 
(Zel'dovich, 1970; Zel'dovich et al. 1982; Doroshkevich, Shandarin, \& Saar 1978) of the cluster 
formation. According to Miller \& Smith (1982), Salvador-Sole \& Solanes (1993), Usami \& Fujimoto 
(1997), in the hierarchical scenario the galaxies could be aligned due to the tidal field of the cluster. 
However, the tidal field of the cluster would apparently be more effective in rich clusters with higher 
mass. 

During the cluster evolution the primordial alignment of galaxies will be altered. First, the alignment 
rate will decrease in the result of mutual interactions between galaxies. Apparently the rate of 
interactions will be higher in rich clusters and especially at the cluster dense central regions. The 
gravitational influence would have smaller effect on the orientation for massive galaxies. Indeed the 
alignment of very massive BCGs (cDs) with the cluster orientation have been found by 
Sastry (1968), Binggeli (1982), Carter \& Metcalfe (1980), Struble (1990), West (1989, 1994), 
Plionis (1994), Fuller, West \& Bridges (1999), Kim et al. (2001), Chambers, Melott, Miller (2002). 

 The inclusion to the cluster content of the faint field galaxies by the secondary infall (Quinn 
\& Binney 1992) (Blanton et al., 2001) with arbitrary orientations will certainly decrease the relative 
number of aligned galaxies. The poorer is the cluster, i.e. the less massive it is, the smaller amount 
of field galaxies would be assembled.  Thus, the primordial alignment is better preserved in poor 
clusters, in which both reasons for altering it, interactions between galaxies and the assembly of 
field galaxies are less effective. 

\section{Acknowledgements}
T-P acknowledges for support through grant DAIP-UGto (0173/19). This research has made use of 
the NASA/IPAC Extragalactic Database (NED), which is operated by the Jet Propulsion Laboratory, 
California Institute of Technology, under contract with the National Aeronautics and Space 
Administration. Funding for SDSS-III has been provided by the Alfred P. Sloan Foundation, the 
Participating Institutions, the National Science Foundation, and the U.S. Department of Energy 
Office of Science. The SDSS-III web site is http://www.sdss3.org/.

SDSS-III is managed by the Astrophysical Research Consortium for the Participating Institutions 
of the SDSS-III Collaboration including the University of Arizona, the Brazilian Participation 
Group, Brookhaven National Laboratory, Carnegie Mellon University, University of Florida, the 
French Participation Group, the German Participation Group, Harvard University, the Instituto 
de Astrofisica de Canarias, the Michigan State/Notre Dame/JINA Participation Group, Johns Hopkins 
University, Lawrence Berkeley National Laboratory, Max Planck Institute for Astrophysics, Max 
Planck Institute for Extraterrestrial Physics, New Mexico State University, New York University, 
Ohio State University, Pennsylvania State University, University of Portsmouth, Princeton University, 
the Spanish Participation Group, University of Tokyo, University of Utah, Vanderbilt University, 
University of Virginia, University of Washington, and Yale University.

\end{document}